\documentclass[12pt]{article}
\usepackage{graphicx}
\usepackage{graphics}
\usepackage{amsthm}
\usepackage{amssymb, amsmath}
\title{Expected exit times of Brownian motion from planar domains:
Complements to a paper of Markowsky} 
\author{Mark W. Coffey\\
Department of Physics\\
Colorado School of Mines\\
Golden, CO  80401\\
(Received $\mbox{~~~~~~~~~~~~~~~~~~~~~~~~~~~~~~~2012}$)}
\date{March 15, 2012}
\begin{document}
\maketitle
\baselineskip=25 pt
\begin{abstract}

We supplement a very recent paper of G. Markowsky concerned with the expected exit
times of Brownian motion from planar domains.  Besides the use of conformal mapping,
we apply results from potential theory.  We treat the case of a wedge-shaped region
exactly, subsuming an example of Markowsky.  A number of other results are presented,
including for a half disc, $n$-grams, and a variety of other regions.

\end{abstract}
 
\medskip
\baselineskip=15pt
\centerline{\bf Key words and phrases}
\medskip 

\noindent

Brownian motion, expected exit time, planar domain, polygon, conformal map, Poisson equation,
half disc, $n$-gram, generalized hypergeometric function, Green function

\vfill
\centerline{\bf 2010 AMS codes} 
60J65, 30C20, 35J05 
 
\baselineskip=25pt
\pagebreak
\medskip
\centerline{\bf Introduction}
\medskip

We will be concerned with developing exact results for the expected exit time of Brownian
motion from certain simply connected planar domains.  The domains include polygons, not
necessarily regular, and others.  The results have applications to other areas of both
pure and applied mathematics.  This Introduction also shows our notation.

Intuitively, the average
exit time of Brownian motion gives a sense of the size of the domain, and the
distance from the starting point to the boundary.  One may then make connections
between such probabilistic considerations and the theory of analytic functions
\cite{burk,davis}.

For this, let {\cal S} be the Schlicht class of analytic functions $f$ on the unit disc
D, normalized such that $f(0)=0$ and $f'(0)=1$.  By the Riemann mapping theorem,
there is an onto map of D to simply connected domains $\Omega$.  In a sense, the
extremal functions in {\cal S} are the identity function $I(z)=z$ (the smallest)
and the Koebe function $K(z)=z/(1-z)^2$ (the largest).  Earlier results \cite{burkholder77,davis} essentially show that the expected exit time of Brownian motion starting at the
origin of a simply connected domain is bounded above and below by the corresponding
times for the Koebe domain and the unit disc, respectively.
For the Koebe domain, the expected exit time is infinite while for the unit disc it
has the value $1/2$.


Let $R$ be an open, connected subset of $\mathbb{R}^n$ with $n \geq 2$, and $X$ a
Brownian motion starting at a point $x$ in $R$.  Then $\tau$ is the first time
$X$ leaves $R$:  $\tau(\omega)=\mbox{inf}\{t>0: X_t(\omega) \notin R\}$.
A result that is convenient for two dimensional applications is provided by the 
following \cite{ban}.
\newline{\bf Lemma 1}.  Let $f(z)=\sum_{n=0}^\infty a_nz^n$ be conformal on D.
Then for Brownian motion starting at $f(0)$, and exit time $\tau$,
$$E_{f(0)}[\tau(f(D))]={1 \over 2}\sum_{n=1}^\infty |a_n|^2.$$
This result can be proved by using the optional stopping theorem for martingales.
It can also be developed by using the Green function for the two dimensional
Laplacian.  

Another approach to finding expected exit times is through potential theory.  Then
the exit time is given by the unique solution of the Poisson equation $\nabla^2 u=-2$
in $R$ with $u$ vanishing on the boundary of the domain (Dirichlet boundary condition)
(e.g., \cite{dynkin}).  Equivalently, if $R$ has a suitably normalized Green function
$G(x,y)$, the expectation of $\tau$ as a function of the starting point $x$ is given by
$E_x[\tau(R)]=\int_R G(x,y)dy$.  For some highly symmetric polygonal domains, 
including the equilateral triangle and the rectangle, there are known eigenfunctions and eigenvalues from which the Green function may be constructed.  As far as the 
equilateral triangle, there is research starting with Lam\'{e} \cite{lame} and ongoing
\cite{alabert,oldham,pinsky1980,pinsky1985,prager}.  Other recent work also suggests how to develop results for other symmetric domains including a regular rhombus and hexagon
\cite{mccartin,mccartin2}.    
For the equilateral triangle, the eigenvalues of the Laplacian with zero Dirichlet
boundary condition are given by $\lambda_{mn}={{16\pi^2} \over {27}}(m^2+n^2-mn)$,
$m,n=0, \pm 1, \ldots$ subject to:  $m+n$ is a multiple of $3$, $m\neq 2n$, and
$n \neq 2m$.  In particular, Pinsky \cite{pinsky1980,pinsky1985} investigated the
multiplicity of the eigenvalues, observing that any eigenvalue can be written as
the norm of an integer in the quadratic field $Q(\sqrt{-3})$, and the symmetry of the eigenfunctions.

Very recently Markowsky \cite{markowsky} investigated expected exit times from
simply connected planar domains.  In particular, an example concerns the expected exit time from a regular polygon $U_m$ of $m$ sides ($m$-gon).  In this circumstance, the well
known mapping from the unit disc to the polygon is provided by a case of the
Schwarz-Christoffel transformation with all interior angles equal, with value $\pi^{(m-2)/m}$.

The expected exit time from $U_m$ is expressed in terms of a generalized 
hypergeometric function $_pF_q$,
$$E_0[\tau(U_m)]={m^2 \over {2B^2(1/m,1-2/m)}} ~_4F_3(1/m,1/m,2/m,2/m;1+1/m,1+1/m,1;1),
\eqno(1.1)$$
where $B(x,y)=\Gamma(x)\Gamma(y)/\Gamma(x+y)$ is the Beta function and $\Gamma$ is the Gamma function.  The generalized hypergeometric function is defined by
\[{}_pF_q(a_1, \ldots ,a_p;b_1, \ldots ,b_q;x)=\sum_{k=0}^{\infty} \frac{(a_1)_k \ldots (a_p)_k}{(b_1)_k \ldots (b_q)_k}\; \frac{x^k}{k!}\quad , \quad |x|<1 \eqno(1.2)\]
where \[(\alpha)_k= \frac{\Gamma(\alpha +k)}{\Gamma(\alpha)}\]
is Pochhammer's symbol.  This function has a branch point at $x=1$.  Under suitable conditions on the parameters, convergence may be obtained on the unit circle. 
Amongst many properties, the function $w(x)= ~_pF_q(\{a_i\};\{b_j\};x)$ satisfies the 
linear ordinary differential equation
$$\left[\theta \prod_{j=1}^q(\theta+b_j-1)-x\prod_{i=1}^p(\theta+a_i)\right]w=0, ~~~~
p \leq q+1,$$
when no $b_j$ is a nonpositive integer.  Here, the differential operator $\theta=x {d \over {dx}}$. 
For the sake of brevity, we refer to several standard sources for information on
the functions $_pF_q$ and other special functions \cite{andrews,bailey,erdelyi,grad,luke,rainville,slater}. 

Another special function that we will use is the two-variable Appell hypergeometric
function $F_1$, given by the series \cite{bailey}
$$F_1(a;b_1,b_2;c;x,y)=\sum_{n=0}^\infty \sum_{m=0}^\infty {{(a)_{n+m}(b_1)_m(b_2)_n} \over
{m!n!(c)_{n+m}}}x^my^n.  \eqno(1.3)$$
This function has the integral representation
$$F_1(a;b_1,b_2;c;x,y)={{\Gamma(c)} \over {\Gamma(a)\Gamma(c-a)}}\int_0^1 u^{a-1}(1-u)^{c-a-1}
(1-ux)^{-b_1}(1-uy)^{-b_2}du,  \eqno(1.4)$$  
and satisfies a system of two partial differential equations of second order each 
(e.g., \cite{grad}, p. 1054).  Some reductions for $F_1$ are known, including
$$F_1(\alpha;\beta,\beta';\gamma;x,1)={{\Gamma(\gamma)\Gamma(\gamma-\alpha-\beta')} \over
{\Gamma(\gamma-\alpha)\Gamma(\gamma-\beta')}} ~_2F_1(\alpha,\beta;\gamma-\beta';x), \eqno(1.5a)$$
and
$$F_1(\alpha;\beta,\beta';\gamma;x,x)= ~_2F_1(\alpha,\beta+\beta';\gamma;x).  \eqno(1.5b)$$

We first present results based upon the use of conformal mapping and Lemma 1 for
wedge-shaped regions, the half disc, symmetric star-shaped regions ($n$-grams), and
a region bounded by circular arcs.  
We then give results for a variety of domains by providing the solution of the Poisson
equation with constant source.


\medskip
\centerline{\bf Results from conformal mapping}
\medskip

The following generalizes Example 5 of \cite{markowsky}.
{\newline \bf Proposition 1}.  Let $\mathbb{H}=\{\mbox{Re} ~z>0\}$ and let $\mathbb{H}^p=\{|\mbox{arg}(z)|<\pi p/2\}$ for $p \leq 1$, where arg$(z)$ takes values in $(-\pi,\pi]$.
Then
$$E_1[\tau(\mathbb{H}^p)]={1 \over 2}\sum_{m=1}^\infty {1 \over m^2}{1 \over {B^2(m,p)}}
 ~_2F_1^2(-m,-p;1-m-p;-1).  \eqno(2.1)$$

Here, $\mathbb{H}$ is the infinite wedge of width $\pi p$ centered on the positive
real axis, and $\mathbb{H}^1=\mathbb{H}$.  This Proposition confirms that \cite{spitzer}
$E_1[\tau(\mathbb{H}^p)]< \infty$ if $p<1/2$, and $E_1[\tau(\mathbb{H}^p)]=\infty$
if $p \geq 1/2$.  

{\it Proof}.  The conformal map from D to $\mathbb{H}^p$ is $g(z)=(1+z)^p/(1-z)^p$.
(As illustrated in Figure 1.)
We develop the Maclaurin expansion of this function with a double application of the
binomial theorem.  In fact, we do this slightly more generally with a function $g_q$:
$$g_q(z)={{(1+z)^q} \over {(1-z)^p}}=(1+z)^q \sum_{n=0}^\infty {{-p} \choose n}(-1)^n z^n
=(1+z)^q \sum_{n=0}^\infty {{(p)_n} \over {n!}}z^n$$
$$=\sum_{j=0}^\infty \sum_{n=0}^\infty {q \choose j}{{(p)_n} \over {n!}}z^{j+n}$$
$$=\sum_{j=0}^\infty \sum_{m=j}^\infty {q \choose j} {{(p)_{m-j}} \over {(m-j)!}}z^m$$
$$=\sum_{m=0}^\infty \sum_{j=0}^m {q \choose j}{{(p)_{m-j}} \over {(m-j)!}}z^m$$
$$=\sum_{m=0}^\infty a_m z^m, \eqno(2.2)$$
where
$$a_m=\sum_{j=0}^m {q \choose j}{{(p)_{m-j}} \over {(m-j)!}}= {{\Gamma(m+p)} \over
{\Gamma(m+1)\Gamma(p)}} ~_2F_1(-m,-q;1-m-p;-1).  \eqno(2.3)$$
Then by applying Lemma 1 with $q=p$ we obtain the Proposition.  For any $p$ or $q$,
the first term of the $_2F_1^2$ factor in (2.1) is $1$.  Then $p=1/2$, where the
factor $\Gamma(1-2p)$ diverges, provides the critical case.  [For further details on
this point, see the special case discussed next.]

We may easily verify that at $q=0$ a case of Markowsky is recovered.  We have
$${1 \over 2}\sum_{m=1}^\infty {1 \over m^2}{1 \over {B^2(m,p)}}
={1 \over {\Gamma^2(p)}}\sum_{m=0}^\infty {1 \over {(m+1)^2}} {{\Gamma^2(m+p+1)} \over
{\Gamma^2(m+1)}}$$
$$={p^2 \over 2}\sum_{m=0}^\infty {1 \over {(m+1)^2}}{{(p+1)_m^2} \over {(m!)^2}}$$
$$={1 \over 2} [~_2F_1(p,p;1;1)-1]={1 \over 2}\left[{{\Gamma(1-2p)} \over {\Gamma^2(1-p)}}-1\right], \eqno(2.4)$$
being equivalent to (3.13) in \cite{markowsky}.  In relation to Proposition 1,
Markowsky \cite{markowsky} wrote that ``the Taylor series expansion for $g$ appears to
be unwieldy for arbitrary $p$".  

\begin{figure}[h]
\begin{center}
\includegraphics[height=3.0in,width=6.0in,angle=0]{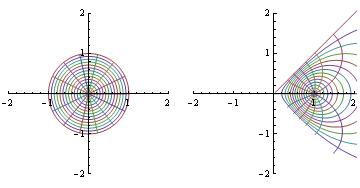} 
\caption{Conformal map from unit disc onto wedge.}
\end{center}
\end{figure}

We next obtain a hypergeometric identity.
\newline{\bf Proposition 2}.  For $0 \leq p <1/2$,
$$\mbox{sec}(\pi p)-1=\sum_{m=1}^\infty {1 \over m^2}{1 \over {B^2(m,p)}}
 ~_2F_1^2(-m,-p;1-m-p;-1).  \eqno(2.5)$$

{\it Proof}.  We solve the Poisson equation for a wedge-shaped region of half angle
$\alpha$.  Using polar coordinates, $\nabla^2 v=-2$ reads
$${1 \over r}{\partial \over {\partial r}}\left(r {{\partial v} \over {\partial r}}\right)
+{1 \over r^2}{{\partial^2 v}\over {\partial \theta^2}}=-2.  \eqno(2.6)$$
We make the solution ansatz $v(r,\theta)=r^n g(\theta)$, with $n$ and $g(\theta)$ to be determined.  Since then $g''(\theta)+n^2g(\theta)=-2r^{2-n}$, we take $n=2$.  With a particular 
solution $g_p(\theta)=-1/2$, the solution for $v$ has the form
$$v(r,\theta)=r^2\left(a \cos 2\theta + b\sin 2\theta-{1 \over 2}\right).  \eqno(2.7)$$
The boundary condition $v(r,\theta=\pm \alpha)=0$ gives $a=1/(2\cos 2\alpha)$ and $b=0$, in
which case
$$v(r,\theta)={r^2 \over 2}\left({{\cos 2\theta} \over {\cos 2\alpha}}-1\right).  \eqno(2.8)$$
The wedge being an unbounded domain, there is the question of the uniqueness of the solution.
As the solution (2.8) grows no faster than $|z|^2$, it is indeed the unique solution
\cite{burk}.

For the point $f(0)=1$, we evaluate $v(r=1,\theta=0)$, putting $2\alpha=\pi p$.  Using
Proposition 1, Proposition 2 then follows. \qed

{\bf Corollary 1}.
$$\sqrt{2}=1+\sum_{m=1}^\infty {1 \over m^2}{1 \over {B^2(m,1/4)}}
 ~_2F_1^2\left(-m,-{1 \over 4};{3 \over 4}-m;-1\right).  \eqno(2.9)$$

{\it Remark}.  The Green function for the wedge can be obtained explicitly via separation of
variables and Fourier series.  We could then integrate over the wedge to determine the 
solution of the Poisson equation.  We have avoided that prospect.

The following gives the expected exit time from the half disc
$D_+=\{|z|<1, \mbox{Im} ~z>0\}$ from the point $f(0)=-i(1-\sqrt{2})$.
\newline{\bf Proposition 3}.
$$E_{f(0)}[\tau(f(D_+))]=2\sum_{m=1}^\infty\left[1-{1 \over \sqrt{2}}\sum_{\ell=0}^{[m/2]}
{{1/2} \choose \ell}\right]^2.  \eqno(2.10)$$

\begin{figure}[h]
\begin{center}
\includegraphics[height=3.0in,width=6.0in,angle=0]{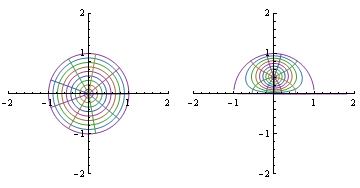} 
\caption{Conformal map from unit disc onto upper half disc.}
\end{center}
\end{figure}

{\it Proof}.  The conformal map of $D_+$ onto the unit disc is given by \cite{olver} 
(Example 7.35, p. 254)
$$\zeta={{z^2+2iz+1} \over {z^2-2iz+1}}.  \eqno(2.11)$$
Therefore, the inverse mapping of the unit disc onto the half disc that we require is
$$\zeta^{-1}(z)={i \over {(z-1)}}[1+z-\sqrt{2}\sqrt{1+z^2}].  \eqno(2.12)$$
This mapping is illustrated in Figure 2.
We then employ binomial expansion to develop the power series of (2.7),
$$\zeta^{-1}(z)=\sum_{m=0}^\infty z^m\left[1+z-\sqrt{2}\sum_{\ell=0}^\infty {{1/2} \choose
\ell} z^{2\ell}\right].  \eqno(2.13)$$
Reordering the sums here and applying Lemma 1 gives the Proposition.  \qed

{\it Remarks}.  The approximate numerical value of (2.10) is $0.191807$.

The other inverse mapping from (2.11),
$$\zeta^{-1}(z)={i \over {(z-1)}}[1+z+\sqrt{2}\sqrt{1+z^2}].  \eqno(2.14)$$
takes the unit disc onto the region outside of the half disc in the lower half plane,
and is illustrated in Figure 3.

\begin{figure}[h]
\begin{center}
\includegraphics[height=3.0in,width=6.0in,angle=0]{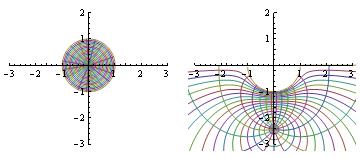} 
\caption{Conformal map from unit disc onto lower half plane exterior to half disc.}
\end{center}
\end{figure}

Results are known for the Green function for the Dirichlet and Neumann problems for a half 
disc \cite{begehr}.  Therefore, it may be possible to complement Proposition 2 with an 
approach of integrating the Dirichlet Green function \cite{begehr} (pp. 257-259) over the half disc.  We note that this method gives in principle a way to find the expected exit time for any point within the half disc.  

In fact, with the reflection symmetry, we may anticipate that a difference of Green 
functions for the circle will suffice.  So for the special homogeneous boundary condition, we have
$$G(z,\zeta)={1 \over {2\pi}}\ln\left|{{(z-\zeta)(1-z\zeta)} \over {(1-z\bar{\zeta})
(z-\bar{\zeta})}}
\right|.  \eqno(2.15)$$
Introducing $z=re^{i\theta}$ and $\zeta=\rho e^{i\phi}$, if the half disc has radius $r_0$,
we have the expression
$$G(z,\zeta)={1 \over {4\pi}}\left[\ln \left({{r^2+\rho^2-2r\rho \cos(\theta-\phi)} \over
{r_0^2+{{r^2\rho^2} \over r_0^2}-2r\rho\cos(\theta-\phi)}}\right)-\ln\left({{r^2+\rho^2-2r\rho \cos(\theta+ \phi)} \over {r_0^2+{{r^2\rho^2} \over r_0^2}-2r\rho\cos(\theta+\phi)}}\right)\right].  \eqno(2.16)$$
This Green function is symmetric in $z$ and $\zeta$, as it should be.
If the half disc has radius $r_0$, then the solution for the expected exit time is given 
by
$$u(r,\theta)=-2 \int_0^{\pi}\int_0^{r_0} G(r,\rho,\theta,\phi)\rho d\rho d\phi. \eqno(2.17)$$
Noting that, for instance,
$$\int_0^{n \pi} \ln(a^2-2ab\cos x+b^2)dx=2\pi n \ln[ \mbox{max}(|a|,|b|)], \eqno(2.18)$$ 
it appears that these integrals can be evaluated explicitly, but are lengthy.  The solution 
for a disc is reviewed in the Appendix.  As well, (2.18) is verified there, and other
integrals are demonstrated.  The latter suffice to perform (2.17) first in terms of dilogarithms, which details will be given elsewhere \cite{coffeydraft}.  Upon combination of all contributions, and simplication, the solution of the Poisson equation can be written in terms of elementary functions.

Indeed, from our exact solution for the Poisson equation and Proposition 3 we have obtained
the following identity.
{\newline \bf Corollary 2}.
$${1 \over 2}\sum_{m=1}^\infty\left(\sum_{\ell=[m/2]+1}^\infty {{1/2} \choose \ell}\right)^2
=\sqrt{2}-1-{1 \over \pi}.$$

Similarly, the Dirichlet Green function for a quarter disc may be written.  We expect that it
too can be integrated to provide an explicit solution of the Poisson equation.

We next consider $n$-grams, non-convex star-shaped regions of $2n$ vertices with alternating interior angles $\pi\alpha_1$ and $\pi\alpha_2$, with $\alpha_1<1<\alpha_2$.  The following result is useful since it implies bounds for the expected exit time.  We let $R$ denote the radius of the circle circumscribing the $n$-gram, and $R_D$ the inradius, the radius of the largest inscribed circle.  
{\newline \bf Proposition 4}.  For an $n$-gram of alternating exterior angles $\pi \mu_1$ and
$\pi \mu_2$,
$$R={{\Gamma(1-\mu_2)\Gamma(1+1/n)} \over {\Gamma(1-\mu_2+1/n)}} ~_2F_1\left(\mu_1,{1 \over n},
1-\mu_2+{1 \over n},-1\right), \eqno(2.19a)$$
and
$$R_D={{\Gamma(1-\mu_1)\Gamma(1+1/n)} \over {\Gamma(1-\mu_1+1/n)}} ~_2F_1\left(\mu_2,{1 \over n},1-\mu_1+{1 \over n},-1\right).  \eqno(2.19b)$$

{\it Proof}.  We let $\pi\mu_i=\pi(1-\alpha_i)$, $i=1$, $2$ be the corresponding alternating
exterior angles.  Because the sum of all exterior angles is $2\pi$, we have
$\mu_1+\mu_2=2/n$.  The Schwarz-Christoffel mapping from the unit disc onto the $n$-gram
has the form 
$$w(z)=A\int_0^z {{\prod_{j \mbox{\tiny{even}}} (z-\omega_j)^{\mu_j}} \over 
{\prod_{j \mbox{\tiny{odd}}} (z-\omega_j)^{\mu_j}}}dz + B, \eqno(2.20)$$
where $A$ and $B$ are constants that may be included to dilate and rotate, and translate,
respectively, the $n$-gram.  We take $\mu_{odd}=\mu_1>0$ and $\mu_{even}=-\mu_2>0$, 
and $\omega_j$ as $2n$th roots of unity, so that
$\prod_{j \mbox{\tiny{even}}}^{2n} (z-\omega_j)=z^n-1$ and $\prod_{j \mbox{\tiny{odd}}}^{2n} (z-\omega_j)=z^n+1$.  Then we may write
$$w(z)=\int_0^z {{(1-z^n)^{-\mu_2}} \over {(1+z^n)^{\mu_1}}}dz.  \eqno(2.21)$$
With a change of variable, 
this integral could be evaluated in terms of $F_1$ using the representation (1.4).
Instead, we perform binomial expansion, integrate term-wise, and appeal to the series
definition (1.3):
$$w(z)=\sum_{j=0}^\infty \sum_{k=0}^\infty (-1)^j {{-\mu_2} \choose j}{{-\mu_1} \choose k}
{z^{n(j+k)+1} \over {[n(j+k)+1]}}$$
$$=z \sum_{j=0}^\infty \sum_{k=0}^\infty (-1)^k {{(\mu_2)_j} \over {j!}} {{(\mu_1)_k} \over {k!}} {{(1/n)_{j+k}} \over {(1+1/n)_{j+k}}} z^{n(j+k)}$$
$$=z F_1\left({1 \over n};\mu_1,\mu_2;1+{1 \over n};z^n;-z^n\right).  \eqno(2.22)$$
To find the radius and inradius, we evaluate at the preimage points $z=\pm 1$, using (1.5).

{\it Remarks}.  A pentagram example is given in \cite{nehari} (pp. 196-197).  

When $\mu_1=1-1/n$ and $\mu_2=3/n-1$, the $_2F_1$ factors in (2.19) may be reduced in terms
of ratios of Gamma functions.  

If we reorder series in (2.22) we may write the Maclaurin expansion for the mapping $w(z)$,
$$w(z)=\sum_{m=0}^\infty \sum_{j=0}^m (-1)^j {{-\mu_2} \choose j}{{-\mu_1} \choose {m-j}}
{z^{nm+1} \over {(nm+1)}}.  \eqno(2.23)$$
Then by Lemma 1 we obtain the following.
{\newline \bf Corollary 3}. For an $n$-gram of alternating exterior angles $\pi \mu_1$ and
$\pi \mu_2$, the expected exit time from the origin is
$$E_0[\tau(f(D))]={1 \over 2}\sum_{k=1}^\infty {1 \over k^2}\left|\sum_{j=0}^{[(k-1)/n]}
(-1)^j {{-\mu_2} \choose j}{{-\mu_1} \choose {{{k-1} \over n}-j}}\right|^2.  \eqno(2.24)$$  

For a univalent map $f$ of the unit disc, there is a number ${\cal U}$, independent of $f$,
such that the inradius $R_{f(D)} \geq {\cal U} |f'(0)|$.  The number ${\cal U}$ is called the
univalent or schlicht Bloch-Landau constant.  Upper and lower bounds are known,
$0.570884 < {\cal U} < 0.6563937$ (e.g., \cite{carrolljoc}), based upon using certain
constructed domains.  For the $n$-gram family, the factor $f'(0)=1$.

For the next exact result, we let $L$ be the region bounded by the circles $C_1$,
$|z-1|=\sqrt{2}$ and $C_2$, $|z+1|=\sqrt{2}$ and containing the origin.  
{\newline \bf Proposition 5}.
$$E_0[\tau(L)]={1 \over \pi}-{1 \over 2}.  \eqno(2.25)$$

{\it Proof}.  We will use the following.
{\newline \bf Lemma 2}.  A conformal map from $L$ to the unit disc $|w|<1$ is
given by $w=2z/(1-z^2)$.  

{\it Proof}.  For points on the circular arc $C_1$, $z=1+\sqrt{2}e^{i\theta}$,
with $3\pi/4 \leq\theta\leq5\pi/4$, so that
$$w={{1+\sqrt{2}e^{i\theta}} \over {-e^{i\theta}(\sqrt{2}+e^{i\theta})}}.$$
This gives
$$|w|=\left|{{1+\sqrt{2}(\cos \theta+i\sin \theta)} \over {\sqrt{2}+\cos \theta+i
\sin \theta}}\right|=\left[{{(1+\sqrt{2}\cos \theta)^2+2\sin^2\theta} \over
{(\sqrt{2}+\cos \theta)^2+\sin^2 \theta}}\right]^{1/2}=1.$$
Similarly, for points on the circular arc $C_2$, $z=-1+\sqrt{2}e^{i\theta}$ with
$-\pi/4\leq \theta \leq \pi/4$.  Then
$$w={{-1+\sqrt{2}e^{i\theta}} \over {e^{i\theta}(\sqrt{2}-e^{i\theta})}},$$
and again $|w|=1$.  

In addition, for $f(z)=2z/(1-z^2)$, $f'(z)=2(1+z^2)/(1-z^2)^2$.  In particular,
$f'$ does not vanish within $L$.  With $f(z)$ being one-to-one in this region,
it follows that $f$ is conformal.  To verify that that the interior of $L$ is
mapped into the interior of the unit disc, we may consider the vertical line
segment $z=it$ with $|t|<1$.  Then $|w|=2|t|/(1+t^2)<1$.  \qed

From the Lemma, the inverse mapping from the unit disc to $L$, illustrated in
Figure 4, is given by
$$z(w)={1 \over w}[-1+\sqrt{1+w^2}],  \eqno(2.26)$$
and its Maclaurin expansion is
$$z(w)=\sum_{\ell=1}^\infty {{1/2} \choose \ell}w^{2\ell-1}=\sum_{\ell=0}^\infty
{{1/2} \choose {\ell+1}}w^{2\ell+1}.  \eqno(2.27)$$
Then by Lemma 1, the expected exit time is given by
$$E_0[\tau(L)]={1 \over 2}\sum_{n=1}^\infty {{1/2} \choose {(n+1)/2}}^2={2 \over \pi}
-{1 \over 2},  \eqno(2.28)$$
the sum being a special case of
$$\sum_{n=1}^\infty {{1/2} \choose {(n+1)/2}}^2 z^n={1 \over z}\left[-1+ ~_2F_1\left(-{1 
\over 2}, -{1 \over 2};1;z^2\right)\right].  \eqno(2.29)$$
\qed

\begin{figure}[h]
\begin{center}
\includegraphics[height=4.0in,width=6.0in,angle=0]{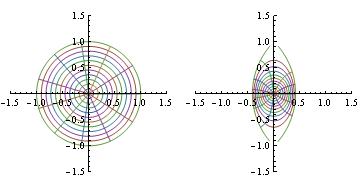} 
\caption{Mapping of unit disc to lens-shaped domain.}
\end{center}
\end{figure}

{\it Remark}.  The mapping (2.26) is a special case of
$$w_p(z)={{[(1+z)^p-(1-z)^p]i} \over {(1+z)^p+(1-z)^p}}.  \eqno(2.30)$$
Here the circular arcs meet at an angle of $\pi p$.
(2.26) then corresponds to $p=1/2$ and an inessential pre-rotation $z \to iz$ of the
unit disc.

\medskip
\centerline{\bf Other results from potential theory}
\medskip

The following solutions of the two-dimensional Poisson equation emphasize that exact
results for average exit time may be found {\it for all} points of certain polygons and
other domains.  Specifically, we consider solutions of $(\partial_x^2+\partial_y^2)u(x,y)=-2$
with zero Dirichlet boundary condition.

{\it Equilateral triangle}.  Let the triangle of side length $a$ with origin at the
center be oriented as in Figure 5.  The solution of the Poisson equation is
$$u(x,y)={1 \over {18a}}(2\sqrt{3}y+a)(\sqrt{3}y+3x-a)(\sqrt{3}y-3x-a).  \eqno(3.1)$$
Should the triangle be inscribed in a circle of unit radius, then by the law of cosines
we have $a=\sqrt{3}$.  Then $u(0,0)=a^2/18=1/6$, in agreement with $E_0[\tau(U_3)]=1/6$
\cite{alabert}.  

Here the area of the triangle is $A=\sqrt{3}a^2/8$, and the product form of the solution
(3.1) may be recognized as parallel to that of \cite{alabert} given in terms of barycentric coordinates $(\lambda_1,\lambda_2,\lambda_3)$, 
$2\sqrt{3}A\lambda_1\lambda_2\lambda_3$.  At least at the time, the authors of
\cite{alabert} were unaware of the exact solution (3.1).  However, the solution (3.1)
has been well known in torsion theory for a long time (e.g., \cite{sokolnikoff}).

\begin{figure}[h]
\begin{center}
\includegraphics[height=4.0in,width=5.50in,angle=0]{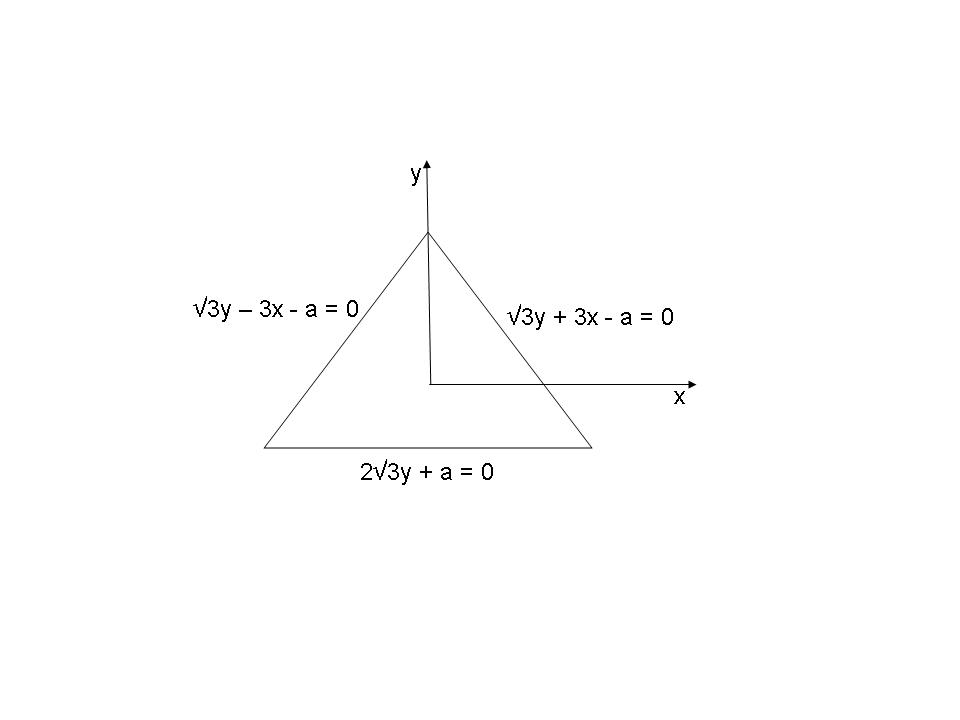} 
\caption{Equilateral triangle domain.}
\end{center}
\end{figure}

{\it A circular domain with circular indentation}.  We next consider the domain of Figure 6
with radii $a$ and $b$ with $a \geq b$.  The solution of the Poisson equation is given by
\cite{sokolnikoff} (pp. 126-127) 
$$u(x,y)=-{1 \over 2}(x^2+y^2-b^2)\left(1-{{2ax} \over {x^2+y^2}}\right), \eqno(3.2)$$
or, in terms of polar coordinates with $r^2=x^2+y^2$ and $\theta=\tan^{-1}(y/x)$,
$$u(r,\theta)=-{1 \over 2}(r^2-b^2)\left(1-2a{{\cos \theta} \over r}\right).  \eqno(3.3)$$
Again we know the expected exit time of Brownian motion from all points of the domain.
A special case is that from the point at the center of the larger circle,
$u(r=a,\theta=0)={1 \over 2}(a^2-b^2)$.  We also obtain the expected result as $b \to 0$.

\begin{figure}[h]
\begin{center}
\includegraphics[height=4.0in,width=5.0in,angle=0]{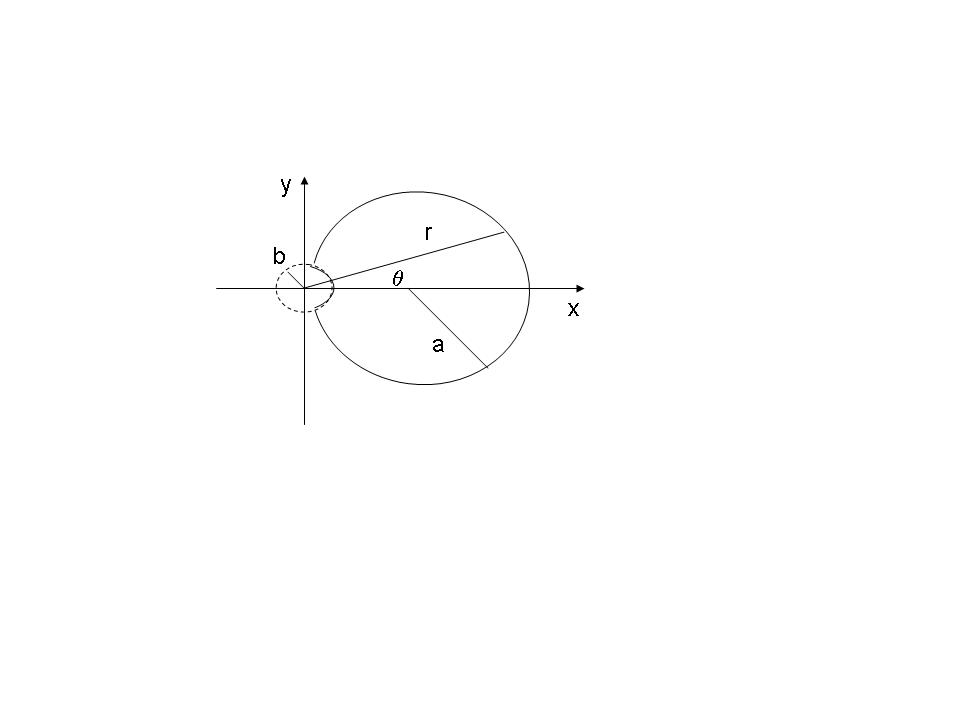} 
\caption{Circular domain with circular cutout.}
\end{center}
\end{figure}

{\it Isosceles right triangle domain}.  The geometry, with two sides of the triangle of 
length $a$ parallel to the coordinate axes, and the hypotenuse along the line $y=-x$, is detailed in Figure 7.  By combining homogeneous and particular solutions,
the solution of the Poisson equation has the form
$$u(x,y)=-\left[{1 \over 2}(x^2+y^2)-\psi(x,y)\right], \eqno(3.4)$$
where $\psi$ is a harmonic function.  We let $k_n=(2n+1)\pi/a$.  Then $\psi(x,y)$ is given
by \cite{sokolnikoff} (pp. 133-134)
$$\psi(x,y)=-xy+{a \over 2}(x+y)-{{4a^2} \over \pi^3}\sum_{n=0}^\infty {{(-1)^n} \over
{(2n+1)^3\sinh(k_na/2)}}(\sinh k_ny\cos k_nx+\sinh k_n x\cos k_ny).  \eqno(3.5)$$
This function has been constructed so that its restriction to the sides of the triangle
is $(x^2+y^2)/2$; this may be verified with a certain cosine Fourier series.  Hence (3.4) satisfies the zero Dirichlet boundary condition.  A special case of (3.4) is for points
along $x=y$, wherein
$$u(x,y=x)=-2x^2+ax-{{8a^2} \over \pi^3}\sum_{n=0}^\infty {{(-1)^n} \over
{(2n+1)^3\sinh(k_na/2)}}\sinh k_nx\cos k_nx.  \eqno(3.6)$$
One may compare the diffusion application in \cite{smith} with $a=2$.

\begin{figure}[h]
\begin{center}
\includegraphics[height=4.0in,width=5.50in,angle=0]{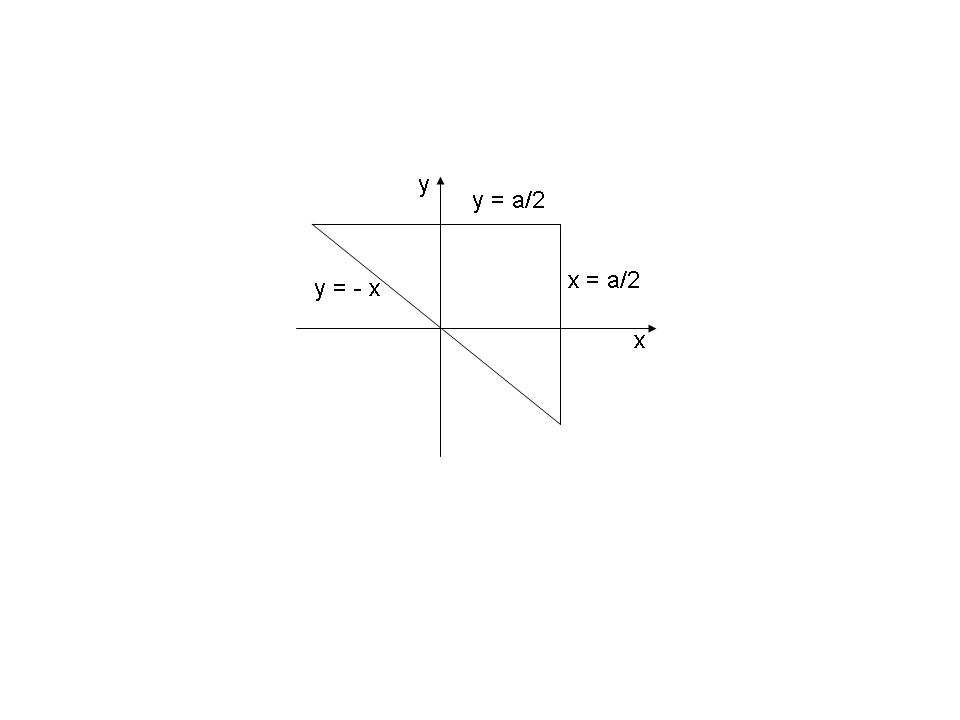} 
\caption{Right isosceles triangle domain.}
\end{center}
\end{figure}

{\it Elliptical domain}.  Let an ellipse be given by the equation $x^2/a^2+y^2/b^2=1$.
Its foci are at $\pm f= \pm \sqrt{a^2-b^2}$ and its eccentricity $e=\sqrt{1-b^2/a^2}$.  
The solution of the Poisson equation with zero Dirichlet boundary condition is
$$w(x,y)={{a^2b^2} \over {a^2+b^2}}\left(1-{x^2 \over a^2}-{y^2 \over b^2}\right).  \eqno(3.7)$$
The maximum average exit time is evident, $w(0,0)=a^2b^2/(a^2+b^2)$.  For $a=b=R$ we recover the known result for a circular disc.  
A conformal mapping of the interior of an ellipse to the unit disc is known, and
involves the elliptic function sn \cite{nehari} (pp. 295-296), \cite{ostrowski}.

{\it Rectangular domain}.  Omitting some details, we
write the solution of the Poisson equation with constant source $-2$ on a $2a \times 2b$ rectangle in the $xy$ plane,
$$u(x,y)=-x^2+a^2-{4 \over a}\sum_{n=0}^\infty {{(-1)^n} \over \alpha_n^3}{{\cosh \alpha_n y} \over {\cosh \alpha_n b}}\cos \alpha_n x, \eqno(3.8)$$
where $\alpha_n \equiv (n+1/2)\pi/a$.  This solution follows from the use of 
separation of variables and the application of Fourier series in the $x$ direction.
Once again there are extensive physical realizations of this solution, including electrostatic
and gravitational potentials, the transverse deflection of a membrane, and steady,
incompressible fluid flow.  
The maximum average exit time occurs at the center of the rectangle,
$$u(0,0)=a^2\left[1-{{32} \over \pi^3}\sum_{n=0}^\infty {{(-1)^n} \over {(2n+1)^3}}
\mbox{sech}\left[\left(n+{1 \over 2}\right){\pi \over a}b\right]\right].$$
From (3.8) we may then supplement relations (3.21) and (3.22) of Markowsky's work
\cite{markowsky},
$$E_0[\tau(U_4)]=~_4F_3(1/4,1/4,1/2,1/2;5/4,5/4,1;1){8 \over {B^2(1/4,1/2)}}$$
$$={{64} \over \pi^4}\sum_{n=1}^\infty \sum_{m=1}^\infty {{(-1)^{m+n}} \over
{(2m-1)(2n-1)[(2m-1)^2+(2n-1)^2]}}, \eqno(3.9)$$
with 
$$E_0[\tau(U_4)]=~_4F_3(1/4,1/4,1/2,1/2;5/4,5/4,1;1){8 \over {B^2(1/4,1/2)}}$$
$$={1 \over 2}-4\sqrt{2}\sum_{n=0}^\infty {{(-1)^n} \over \alpha_n^3}
{1 \over {\cosh \alpha_n/\sqrt{2}}}, \eqno(3.10)$$
for the expected exit time from the square.  For the square inscribed within the unit
disc, we have put $a=b=1/\sqrt{2}$.  It may be noted that we thus have a form
of the expected exit time for a rectangle of arbitrary dimensions and for an arbitrary
starting point.  

The double series form, (3.9) above, (3.22) of \cite{markowsky}, is based upon the double Fourier sine series solution in \cite{knight} (p. 71).  It is then apparent that that
representation and (3.10) are related via the partial fractions forms
$$\tan\left({{\pi x} \over 2}\right)={{4x} \over \pi}\sum_{k=1}^\infty {1 \over 
{[(2k-1)^2-x^2]}}, \eqno(3.11a)$$
$$\cot\left({{\pi x} \over 2}\right)={1 \over {\pi x}}+{{2x} \over \pi}\sum_{k=1}^\infty {1 \over {(x^2-k^2)}}, \eqno(3.11b)$$
and 
$$\sec(ix)=\mbox{sech} ~x={4 \over \pi}\sum_{k=0}^\infty {{(-1)^k (2k+1)} \over
{[(2k+1)^2 +x^2]}}.  \eqno(3.11c)$$
Indeed, from (3.11a) and (3.11b) we have
$$\sum_{m=1}^\infty {{(-1)^m} \over {(2m-1)}}{1 \over {[(2m-1)^2+x^2]}}={\pi \over 8}
{1 \over x^2}\left[-2+\cot[(\pi/4)(ix+1)]+\tan[(\pi/4)(ix+1)] \right].  \eqno(3.12)$$

For an infinite strip of width $2a$, we have the special case of (3.8) $u(x)=a^2-x^2$.

\medskip
\centerline{\bf Acknowledgements}
I thank W. McCollom for useful discussions, and G. Markowsky for reading the manuscript.

\pagebreak
\centerline{\bf Appendix:  Green function solution of the Poisson equation for a disc and
integration results}
\medskip

Here we first use the Green function corresponding to the first term on the right side of (2.16)
to solve the two dimensional Poisson equation with homogeneous Dirichlet boundary condition.
Noting that by symmetry, the solution can not depend upon $\theta$, we have
$$u(r)=-{1 \over {2\pi}}\int_0^{r_0}\int_0^{2\pi}\ln \left({{r^2+\rho^2-2r\rho \cos \phi} \over
{r_0^2+{{r^2\rho^2} \over r_0^2}-2r\rho\cos \phi}}\right)\rho d\rho d\phi.  \eqno(A.1)$$
Using (2.18), one contribution is
$$\int_0^{r_0}\int_0^{2\pi}\ln (r^2+\rho^2-2r\rho \cos \phi)\rho d\rho d\phi
=4\pi \int_0^{r_0}\ln[\mbox{max}(\rho,r)]\rho d\rho$$
$$=4\pi \left[\ln r \int_0^r \rho d\rho +\int_r^{r_0} \rho \ln \rho ~d\rho\right]$$
$$=\pi [2r_0^2 \ln r_0 +r^2-r_0^2].  \eqno(A.2)$$
The other contribution is
$$\int_0^{r_0}\int_0^{2\pi}\ln \left(r_0^2+{{r^2\rho^2} \over r_0^2}-2r\rho\cos \phi \right)\rho d\rho d\phi = 4\pi \int_0^{r_0} \ln \left[\mbox{max}\left(r_0,{{r \rho} \over r_0}\right)\right]
\rho d\rho$$
$$=4\pi \ln r_0 \int_0^{r_0} \rho d\rho=2\pi r_0^2\ln r_0.  \eqno(A.3)$$
By combining (A.2) and (A.3) in (A.1), we recover
$$u(r)={1 \over 2}(r_0^2-r^2).  \eqno(A.4)$$
\qed

{\bf Lemma 3}.  We have the expansion
$$\ln(r^2+\rho^2-2r\rho\cos(\theta \mp \phi))=2\ln \mbox{max}(r,\rho)- 2\sum_{k=1}^\infty {1 \over k}\left({{\mbox{min}(r,\rho)} \over {\mbox{max}(r,\rho)}}\right)^k[\cos k\theta \cos k
\phi \pm \sin k\theta \sin k\phi].  \eqno(A.5)$$

{\it Proof}.  Assume $r>\rho$.  Then
$$\ln(r^2+\rho^2-2r\rho\cos(\theta \mp \phi))=\ln r^2+\ln\left(1+{\rho \over r}-2{\rho \over r}
\cos(\theta \mp \phi)\right)$$
$$=2\ln r+\ln\left[\left(1-{\rho \over r}e^{i(\theta \mp \phi)}\right)\left(1-{\rho \over r}e^{-i(\theta \mp \phi)}\right)\right]$$
$$=2\ln r - 2\sum_{k=1}^\infty {1 \over k} \left({\rho \over r}\right)^k \cos[k(\theta\mp
\phi)], \eqno(A.6)$$
wherein the logarithms were expanded for $\rho/r<1$.  If instead $r<\rho$, we may simply
interchange them in (A.6).  \qed

Relation (2.18) is an immediate special case integral from Lemma 3, putting $\theta=0$.

More generally we have the following, putting Li$_2(z)=\sum_{n=1}^\infty z^n/n^2=-\int_0^z
[\ln(1-t)/t]dt$ (e.g., \cite{andrews}).  This dilogarithm function has a branch point at $z=1$ and can be analytically continued through out the complex plane.  
{\newline \bf Proposition 6}.  For $r>\rho$,
$$\int_0^\pi \ln(r^2+\rho^2-2r\rho\cos(\theta \mp \phi))d\phi$$
$$=2\pi \ln r\pm i\left[\mbox{Li}_2
\left(-e^{-i\theta} {\rho \over r}\right)-\mbox{Li}_2\left(e^{-i\theta} {\rho \over r}\right) 
-\mbox{Li}_2\left(-e^{i\theta} {\rho \over r}\right)+\mbox{Li}_2\left(e^{-i\theta} {\rho \over r}\right)\right].  \eqno(A.7)$$
{\it Proof}.  Using Lemma 3, we have
$$\int_0^\pi \ln(r^2+\rho^2-2r\rho\cos(\theta \mp \phi))d\phi=2\pi \ln r\mp 2\sum_{k=1}^\infty
{{[1-(-1)^k]} \over k^2}\left({\rho \over r}\right)^k \sin k\theta.  \eqno(A.8)$$
Then using the series definition of Li$_2$ gives the Proposition.  \qed

In connection with (2.17), we have the integral
$$\int \mbox{Li}_2(c\rho)\rho d\rho={1 \over {8c^2}}[4c^2\rho^2 \mbox{Li}_2(c\rho)+2(c^2\rho^2
-1)\ln(1-c\rho)-c\rho(2+c\rho)].  \eqno(A.9)$$

This follows from
$$\int \mbox{Li}_2(c\rho)\rho d\rho=\int \sum_{n=1}^\infty {c^n \over n^2}\rho^{n+1} d\rho
=\sum_{n=1}^\infty {c^n \over n^2}{\rho^{n+2} \over {(n+2)}}$$
$$={1 \over 2}\sum_{n=1}^\infty c^n\left[{1 \over n^2}-{1 \over {2n}}+{1 \over {2(n+2)}}\right]
\rho^{n+2}.  \eqno(A.10)$$

We also record the following result that is useful when $r<\rho$.
{\newline \bf Lemma 4}.  
$$\int \rho \mbox{Li}_2\left(a{r \over \rho}\right)d\rho
={1 \over 4}\left\{2\rho^2 \mbox{Li}_2\left({{ar} \over \rho}\right)+ar[\rho+ar\ln(\rho-ar)]
-\rho^2\ln\left(1-{{ar} \over \rho}\right) \right\}.  \eqno(A.11)$$

{\it Proof}.  We use three integrations by parts:
$$\int \rho \mbox{Li}_2\left(a{r \over \rho}\right)d\rho=-{1 \over 2}\int \rho\ln(1-ar/\rho)d
\rho+{\rho^2 \over 2} \mbox{Li}_2\left({{ar} \over \rho}\right)$$
$$={1 \over 4}\int{{ar\rho} \over {(\rho-ar)}}d\rho-{\rho^2 \over 4}\ln(1-ar/\rho)+{\rho^2 \over 2} \mbox{Li}_2\left({{ar} \over \rho}\right)$$
$$={{ar\rho} \over 4}+{{a^2r^2} \over 4}\ln(\rho-ar)-{\rho^2 \over 4}\ln(1-ar/\rho)+{\rho^2 
\over 2} \mbox{Li}_2\left({{ar} \over \rho}\right).$$
\qed

\pagebreak

\end{document}